\documentclass[11pt]{article}
\usepackage{amsmath,amsthm}
\usepackage{amssymb}
\usepackage[dvipdfmx]{graphicx,color}
\allowdisplaybreaks
\nopagebreak[3]
\newcommand{\newsection}[1]
{\section{#1}\setcounter{theorem}{0} \setcounter{equation}{0} \par\noindent}

\newtheorem{theorem}{Theorem}

\newtheorem{lemma}[theorem]{Lemma}
\newtheorem{corollary}[theorem]{Corollary}

\newtheorem{definition}[theorem]{Definition}
\newtheorem{remark}[theorem]{Remark}

\newcommand{\beq}{ \begin{equation} }
\newcommand{\eeq}{ \end{equation} }

\newcommand{\br}{{\mathbb R}}
\newcommand{\bc}{{\mathbb C}}

\newcommand{\supp}{\mbox{\rm supp}\ }

\newcommand{\brn}{ { \mathbb{R}^n } }
\renewcommand{\Re}{\operatorname{Re}}

\newcommand{\two}{I\hspace{-1mm}I}
\newcommand{\three}{I\hspace{-1mm}I\hspace{-1mm}I}
\newcommand{\four}{I\hspace{-1mm}V}

\title{
Nonexistence of global weak solutions 
\\
of Klein-Gordon equations 
\\ 
with gauge variant semilinear terms 
\\
in Friedmann-Lema\^itre-Robertson-Walker spacetimes 
}
\author
{
Makoto NAKAMURA
\thanks
{Graduate School of Information Science and Technology,  
Osaka University, 
1-5 Yamadaoka, Suita, Osaka 565-0871, JAPAN.  
E-mail:  \texttt{makoto.nakamura.ist@osaka-u.ac.jp} }
\ \ and \ 
Takuma YOSHIZUMI
\thanks
{Graduate School of Information Science and Technology,  
Osaka University, 
1-5 Yamadaoka, Suita, Osaka 565-0871, JAPAN.  
E-mail:  \texttt{yoshizumi.takuma@ist.osaka-u.ac.jp} }
}

\date{}

\begin{document}

\maketitle

\begin{abstract}
Nonexistence of global weak solutions of Klein-Gordon equations with gauge variant semilinear terms are considered in Friedmann-Lema\^itre-Robertson-Walker spacetimes. 
Effects of spatial expansion or contraction on the solutions are studied through the scale-function and the curved mass.
\end{abstract}

\noindent
{\it Mathematics Subject Classification (2020)}: 
Primary 35L05; Secondary 35L71, 35Q75. 
%35G25 Initial value problems for nonlinear higher-order PDEs
%35L05 Wave equation
%35L15 Initial value problems for second-order hyperbolic equations
%35L70 Second-order nonlinear hyperbolic equations
%35L71 Second-order semilinear hyperbolic equations
%35L72 Second-order quasilinear hyperbolic equations
%35Q60 PDEs in connection with optics and electromagnetic theory
%35Q61 Maxwell equations
%35Q30 Navier-Stokes equations
%35Q75 PDEs in connection with relativity and gravitational theory
%74B05 Classical linear elasticity
%35G20 Nonlinear higher-order equations
%35Q76 Einstein equations

\vspace{5pt}

\noindent
{\it Keywords} : 
semilinear Klein-Gordon equation, Cauchy problem, 
Friedmann-Lema\^itre-Robertson-Walker spacetime

%%%%%%%%%%%%%%%%%%%%%%%%%%%%%%%
%
%%%%%%%%%%%%%%%%%%%%%%%%%%%%%%%

\section{Introduction}

We show the nonexistence of global weak solutions 
of Klein-Gordon equations with gauge-variant semilinear terms in Friedmann-Lema\^itre-Robertson-Walker spacetimes 
(FLRW spacetimes for short).
FLRW spacetimes are solutions of the Einstein equations 
with the cosmological constant under the cosmological principle.
They describe the spatial expansion or contraction, 
and yield some important models of the universe.
Let $n\ge1$ be the spatial dimension, 
$a(\cdot)>0$ be a scale-function defined on an interval $[0,T_0)$ for some $0<T_0\le\infty$,
and let $c>0$ be the speed of light.
The metrics $\{g_{\alpha\beta}\}$ of FLRW spacetimes are expressed by 
$\sum_{0\le \alpha,\beta\le n}g_{\alpha\beta}dx^\alpha dx^\beta
:=
-c^2(dt)^2+
a^2(t)
\sum_{j=1}^n (dx^j)^2$,
where we have put the spatial curvature as zero,  
and $x^0=t$ is the time-variable 
(see e.g., 
\cite{Carroll-2004-Addison, DInverno-1992-Oxford}).
When $a$ is a positive constant, 
the spacetime 
reduces to the Minkowski spacetime.

%%%%%%%%%%%%%%%%%%%%%%%
% Derivation of equations
%%%%%%%%%%%%%%%%%%%%%%%
We denote the first and second derivatives of one variable function $a$ by $\dot{a}$ and $\ddot{a}$.
The Klein-Gordon equation generated by the above metric 
$(g_{\alpha\beta})_{0\le \alpha,\beta\le n}$ is given by  
$(\sqrt{|g|})^{-1}\partial_\alpha\left(\sqrt{|g|}g^{\alpha\beta}\partial_\beta v\right)=m^2v+f(v)$ 
for the determinant $g:=\mbox{det}\left(g_{\alpha\beta}\right)$ 
and the inverse matrix $\left(g^{\alpha\beta}\right)$, 
i.e.,
\beq
\label{Cauchy-0}
\frac{1}{c^2}\partial_t^2v+\frac{n\dot{a}}{c^2a}\partial_tv-\frac{1}{a^2} \Delta v 
+
m^2
%\left(\frac{mc}{\hbar}\right)^2
v+f(v)
=0,
\eeq
where 
$m$ denotes the mass,  
$\Delta:=\sum_{j=1}^n \partial_j^2$ denotes the Laplacian, 
and $f$ denotes a force term.
This equation is rewritten as 
\beq
\label{Eq-KGLambda}
c^{-2}\partial_t^2u-a^{-2}\Delta u
+M^2u
+a^{n/2}f(a^{-n/2}u)=0
\eeq
by the transformation $u:=va^{n/2}$,
where the function $M$ is called curved mass defined by  
\beq
\label{Def-M}
M^2
=
m^2-\frac{n(n-2)}{4c^2}\left( \frac{\dot{a}}{a} \right)^2
-\frac{n}{2c^2}\cdot\frac{\ddot{a}}{a},
\ \ \ \ 
M:=\sqrt {M^2}.
\eeq

In this paper, we consider the Cauchy problem in the gauge-variant case $f(v)=\lambda |v|^p$ in 
\eqref{Eq-KGLambda}, namely,    
\beq
\label{Cauchy}
\left\{
\begin{array}{l}
(c^{-2}\partial_t^2-a^{-2}(t) \Delta+M^2(t))u(t,x)-\lambda a^{-n(p-1)/2}(t)|u|^p(t,x)=0,
\\
u(0,\cdot)=u_0(\cdot),\ \ \partial_tu(0,\cdot)=u_1(\cdot)
\end{array}
\right.
\eeq
for $(t,x)\in [0,T)\times\br^n$ and $0<T\le T_0$, 
where 
$u_0$ and $u_1$ are given initial data.
We expect some dissipative effects when $\dot{a}>0$ in \eqref{Cauchy-0} 
from the term $n\dot{a}\partial_t v/c^2a$.
Actually, the spatial expansion yields dissipative effects in energy estimates for  the Klein-Gordon equation 
(see 
\cite{Nakamura-2014-JMAA, Nakamura-Yoshizumi-2025-JDE}).
The Cauchy problem of \eqref{Cauchy} including gauge invariant semilinear term is widely studied especially on local and global solutions, and their asymptotic behaviors 
(see 
\cite{Baskin-2012-AHP,
Galstian-Yagdjian-2015-NA-TMA, 
Hintz-Vasy-2015-AnalysisPDE,
Nakamura-2021-JMP,
Nakamura-Yoshizumi-2025-JDE,
Yagdjian-2012-JMAA} for closely related results).

Blowing-up solutions for \eqref{Cauchy} have been considered in 
\cite[Proposition 1]{Nakamura-2021-JMP} and 
\cite[Theorems 1.1 and 1.2]{Yagdjian-2009-DCDS} 
in the de Sitter spacetime.
The aim of this paper is to extend these results into general FLRW spacetimes including concrete examples of scale functions of Big-Rip and Big-Crunch.
We consider the real and purely imaginary mass $m\in \br\cup i\br$ 
since $m\in i\br$ plays an important role when we consider the breakdown of symmetry of potential well, or in $\phi^4$-theory in physics 
(see \cite{Nakamura-2021-JMP}).
We refer to \cite{Wei-Yong-2024-JMP} and the references therein 
on blowing-up solutions for small initial data in FLRW spacetimes 
for gauge-variant semilinear terms and vanishing mass $m=0$, i.e., wave equation, 
under some conditions on $p$ concerned with the Strauss conjecture 
(see also \cite{Galstian-Yagdjian-2020-RMP} in the Einstein-de Sitter spacetime).
Our results are for relatively large data and arbitrary $m\in \br\cup i\br$ and $1<p<\infty$.

The case of the gauge invariant semilinear term of the form 
$f(u)=\lambda |v|^{p-1}v$ in \eqref{Eq-KGLambda} 
has been considered by McCollum, Mwamba and Oliver in 
\cite[Theorem 1]{McCollum-Mwamba-Oliver-2024-NA} for $\dot{a}>0$, 
and in \cite[Theorem 1.4]{Nakamura-Yoshizumi-2025-JDE} for $\dot{a}\le 0$ by different techniques from this paper to show the blowing-up of the norm $\|u\|_2$,
while the nonexistence of global weak solutions 
is considered through the blowing-up of $\int_\brn u(t,x)dx$ in this paper.

\vspace{9pt}

\begin{definition}[Weak global solution]
Let $a\in C([0,\infty),\br)$, $M\in C([0,\infty), \bc)$ 
and $h\in L^1_{\mbox{\rm loc}}([0,\infty)\times\brn, \bc)$.
We say that $u\in L^1_{\mbox{\rm loc}}([0,\infty)\times\brn)$ is a weak global solution of the Cauchy problem 
\[
\begin{cases}
c^{-2}\partial_t^2 u-a^{-2}\Delta u+M^2 u+h=0,
\\
u(0,\cdot)=u_0(\cdot),\ \ \partial_t u(0,\cdot)=u_1(\cdot)
\end{cases}
\]
if $u$ satisfies the integral equation 
\begin{multline}
\label{Weak-Sol-Psi}
c^{-2}\int_\brn u_0(x)\partial_t\psi(0,x) dx
-
c^{-2}\int_\brn u_1(x)\psi(0,x)dx
\\
+
\int_0^\infty \int_\brn u(t,x)
\left\{
c^{-2} \partial_t^2\psi(t,x)-a^{-2}(t)\Delta \psi(t,x)+M^2(t)\psi(t,x)
\right\}
 dxdt
\\
+
\int_0^\infty \int_\brn h(t,x)\psi(t,x)dxdt=0
\end{multline}
for any $\psi\in C_0^2([0,\infty)\times \brn)$.
\end{definition}

Throughout the paper, we assume that the solution $u$ of \eqref{Cauchy} satisfies the finite speed of propagation.
Namely, the support of $u$ is inside of the characteristic curve such that 
$\{(t,x); 0\le t<T,\ |x|\le r(t)\}$ 
if the initial data with their supports satisfying 
$\supp u_0\, \cup\, \supp u_1\subset \{x\in\brn; \, |x|\le r_0\}$,
where 
\beq
\label{Def-r}
r(t):=r_0+\int_0^t \frac{c}{a(s)} ds
\eeq
(see \cite[Appendix]{Nakamura-Yoshizumi-2025-JMP} for details).

\begin{theorem}[Nonexistence of global weak solutions]
\label{Thm-2}
Let $n\ge1$, $\lambda>0$, $1<p<\infty$ and $r_0>0$.
Let 
\beq
\label{Data}
u_0\in L^1(\brn), 
\ \ u_1\in L^1(\brn), 
\ \ \supp u_0\cup \supp u_1\subset \{x\in\brn; \, |x|\le r_0\}.
\eeq
Put 
\beq
\label{Def-w0-w1}
w_0:=\Re \int_\brn u_0(x) dx, 
\ \ 
w_1:=\Re \int_\brn u_1(x) dx.
\eeq
Let $a\in C([0,\infty),(0,\infty))$, $M\in C([0,\infty),i\br)$, 
and let $N$ be a number with 
\beq
\label{Thm-2-Assume-aMN}
\inf_{0<t<\infty} M^2(t)>-\infty, 
\ \ 
0\le N<\infty,
\ \ N^2+\inf_{0<t<\infty} M^2(t)\ge0.
\eeq
Let $0<\theta<1$.
Put   
\beq
\label{Def-b-S}
b:=\frac{\lambda}{\omega_n^{p-1} \left(ar^2\right)^{n(p-1)/2} },
\ \ 
S:=\sup_{0<t<\infty}
\left\{
e^{-cNt}
\left(
\frac{N^2+M^2(t)}{(1-\theta)b(t)}
\right)^{1/(p-1)}
\right\},
\eeq
where $\omega_n$ denotes the volume of unit ball in $\brn$,
and assume 
\beq
\label{Assume-S-w}
S<\infty,
\ \ 
w_0>S,
\ \ 
w_1\ge cNw_0,
\eeq
\beq
\label{Thm-2-Assume-1}
\lim_{R\to \infty}
R^{-2}
\left\{
\int_{R/2}^R \left(\min\{R,r(t)\}\right)^n a^{n/2}(t) dt
\right\}^{1/p'}
=0,
\eeq
\beq
\label{Thm-2-Assume-2}
\lim_{R\to \infty}
R^{-2}
\left\{
\int_0^R
\int_{\substack{R/2<|x|<R \\ |x|<r(t)} }
dx \, a^{n/2-2p'}(t) dt
\right\}^{1/p'}=0,
\eeq
\beq
\int_0^\infty\int_\brn |M^2(t)u(t,x)| dxdt<\infty.
\eeq
Then the nontrivial weak solution $u$ of \eqref{Cauchy} does not exist.
\end{theorem}

\vspace{9pt}

%%%%%%%%%%%
%  Example  %
%%%%%%%%%%%

Now, we introduce some concrete examples of the scale-function $a$ and the curved mass $M$.
For $\sigma\in\mathbb{R}$ and the Hubble constant $H\in \br$, 
we put
\begin{equation}
\label{R-Def-T_0}
T_0:= 
\begin{cases}
\infty & \mbox{if}\ \ (1+\sigma)H\ge0,
\\
-\frac{2}{n(1+\sigma)H}(>0) & \mbox{if}\ \ (1+\sigma)H<0,
\end{cases}
\end{equation}
and define $a(\cdot)$ by 
\begin{equation}
\label{Def-a}
a(t):=
\begin{cases}
a_0\exp(Ht) & \mbox{if}\ \ \sigma=-1,
\\
a_0\left\{1+\frac{n(1+\sigma)Ht}{2}\right\}^{2/n(1+\sigma)} & \mbox{if}\ \ \sigma\ne-1
\end{cases}
\end{equation}
for $0\le t<T_0$. 
We note $a_0=a(0)$ and $H=\dot{a}(0)/a(0)$.
This scale-function $a(\cdot)$ describes the Minkowski space when $H=0$ 
(namely, $a(\cdot)$ is the constant $a_0$),
the expanding space when $H>0$ with $\sigma\ge -1$,
the blowing-up space when $H>0$ with $\sigma< -1$ 
(the ``Big-Rip'' in cosmology), 
the contracting space when $H<0$ with $\sigma\le -1$, 
and the vanishing space when $H<0$ with $\sigma> -1$ 
(the ``Big-Crunch'' in cosmology).
It describes the de Sitter spacetime when $\sigma=-1$.
The corresponding curved mass $M$ defined by \eqref{Def-M} is rewritten as  
\beq
\label{M-FLRW}
M^2
=
m^2+\sigma\left(\frac{nH}{2c}\right)^2\cdot 
\left\{1+\frac{n(1+\sigma)Ht}{2}\right\}^{-2}
\eeq
(see (2) in Lemma \ref{Lem-11}, below).
We say that the solution $u$ of \eqref{Cauchy} is a global solution if $u$ exists on the time-interval $[0,T_0)$ 
since $T_0$ is the end of the spacetime.

We note that the squared curved mass $M^2$ changes its signature at $T_1$ 
when $(1+\sigma)H<0$, $\sigma<0$, $m>\sqrt{|\sigma|}n|H|/2c$,
where 
\beq
\label{Def-T1}
T_1:=
-\frac{2}{n(1+\sigma)H}\left( 1-\frac{ \sqrt{|\sigma|}n|H| }{2cm} \right). 
\eeq
We also consider the case $M^2< 0$, 
i.e., purely imaginary curved mass $M\in i\br$,  
since it naturally appears in FLRW spacetimes.

\vspace{10pt}

We obtain the following result from Theorem \ref{Thm-2} for the example of $a$ in \eqref{Def-a}.

\begin{corollary}
[Nonexistence of global weak solutions under \eqref{Def-a}]
\label{Cor-3}
Let $n\ge1$, $\lambda>0$, $1<p<\infty$ and $r_0>0$.
Assume \eqref{Data} and \eqref{Def-w0-w1}.
Let $a$ and $M$ be the functions defined by \eqref{Def-a} and \eqref{Def-M}. 
Let $0<\theta<1$.
Assume \eqref{Assume-S-w} for $S$ and $b$ defined by \eqref{Def-b-S}.
Then the nontrivial weak solution $u$ of \eqref{Cauchy} does not exist under one of the following conditions from (1) to (3).

(1) 
$H=0$, $\sigma\in \br$, $m\in i\br$, $N:=|m|$,
\beq
1<p<
\begin{cases}
\infty & \mbox{if}\ n=1,
\\
1+\frac{2}{n-1} & \mbox{if}\ n\ge2.
\end{cases}
\eeq

(2) 
$H>0$, $-1<\sigma<\infty$,  
\beq
\begin{cases}
m\in i\br, 
\ |m|\ge \frac{\sqrt{\sigma} nH}{2c}, 
\ N:=|m| & \mbox{if}\ \ \sigma>0,
\\
m\in i\br, \ N:=|m| & \mbox{if}\ \ \sigma=0,
\\
m\in i\br, 
\ N:=\sqrt{-m^2-\sigma\left(\frac{nH}{2c}\right)^2} & \mbox{if}\ \ -1<\sigma<0,
\end{cases}
\eeq
\beq
\label{Cor-3-2-p}
1<p<
\begin{cases}
\infty & \mbox{if}\ \ n=1,\ \sigma\ge0,
\\
\infty & \mbox{if}\ \ n=2,\ \sigma=-1+\frac{2}{n},
\\
\frac{(n+1)(1+\sigma)-1}{(n-1)(1+\sigma)-1} 
& \mbox{if}\ n\ge2,\ -1+\frac{2}{n}<\sigma<\infty,
\\
\frac{n+2}{n-2} 
& \mbox{if}\ \ n\ge3,\ \sigma=-1+\frac{2}{n},
\\
-\frac{2+\sigma}{\sigma} 
& \mbox{if}\ \ n=1,\ -1<\sigma<0,
\\
-\frac{2+\sigma}{\sigma} 
& \mbox{if}\ \ n\ge2,\ -1<\sigma<-1+\frac{2}{n}.
\end{cases}
\eeq

(3) 
$H<0$, $\sigma<-1-2/n$, $m\in i\br$, 
$N:=\sqrt{-m^2-\sigma(nH/2c)^2}$,
\beq
\label{p0}
1<p<p_0:=\frac{n\left\{(n+1)(\sigma+1)+1\right\}}{n(n-1)(\sigma+1)+n-4}.
\eeq
\end{corollary}

\begin{remark}
We note that $p_0$ in \eqref{p0} satisfies 
$1<p_0<(n+1)/(n-1)$ for $n\ge2$, 
$p_0=1$ if $\sigma=-1+2/n$.
Moreover, 
$p_0\to(n+1)/(n-1)$ for $n\ge2$ and 
$p_0\to\infty$ for $n=1$ as $\sigma\to-\infty$. 
\end{remark}

\begin{remark}
\label{Rem-5}
We remark on the difference between the results in Corollary \ref{Cor-3} and the results in 
\cite[Corollary 1.2]{Nakamura-Yoshizumi-2025-JMP}.
In \cite{Nakamura-Yoshizumi-2025-JMP},
we have shown blowing-up solutions for $m\in \br\cup i\br$, $1<p<\infty$ in the four cases given by  
(i) $H=0$ with $\sigma\in \br$, 
(ii) $H>0$ with $\sigma\ge-1$,
(iii) $H<0$ with $\sigma\ge0$,
(iv) $H<0$ with $\sigma\le -1$.
Namely, we could consider the real mass $m\in \br$, the power $p$ without upper bound, 
and the case $H>0$ with $\sigma=-1$, and the case $H<0$ with $-1-2/n\le \sigma\le -1$, 
and the case $H<0$ with $\sigma\ge0$, 
which are out of scope of Corollary \ref{Cor-3}.
However, 
\cite{Nakamura-Yoshizumi-2025-JMP} requires 
\beq
\label{Rem-5-100}
w_1\ge 
\max\left\{
cNw_0, 
\sqrt{\frac{2\lambda c^2\theta}{p+1}}
\cdot
\frac{w_0^{(p+1)/2}}{(\omega_n^{2/n}a_0r_0^2)^{n(p-1)/4} }
\right\},
\eeq
and 
\beq
\label{Rem-5-200}
w_0>\sup_{0<t<T_0}
\frac{\omega_n}{e^{cNt}}
\max\left\{
\left(a_0r_0^2\right)^{n/2},
\left(ar^2\right)^{n/2}
\right\}
\cdot
\left\{
\frac{N^2+M^2}{(1-\theta)\lambda}
\right\}^{1/(p-1)},
\eeq
moreover, $r_0\le 2c/a_0|H|$ in the case $H<0$ with $\sigma\le -1$.
On the other hand, Corollary \ref{Cor-3} requires weaker conditions 
$w_1\ge cNw_0$,
\beq
\label{Rem-5-300}
w_0>\sup_{0<t<T_0}
\frac{\omega_n}{e^{cNt}}
\left(ar^2\right)^{n/2}
\cdot
\left\{
\frac{N^2+M^2}{(1-\theta)\lambda}
\right\}^{1/(p-1)},
\eeq
without any condition on $r_0>0$.
\end{remark}

\newsection{Proof of Theorem \ref{Thm-2}}
\label{Sec-Thm-22}
We prove Theorem \ref{Thm-2} in this section.
We use the following fundamental result on the positivity of the solution of an ordinary differential equation of second order.
 
\begin{lemma}(See \cite[Lemma 2.2]{Nakamura-Yoshizumi-2025-JMP}.)
\label{Lem-21}
Let $0<T_0\le \infty$.
Let $M^2\in C([0,T_0),\br)$, $b\in C([0,T_0),(0,\infty))$.
Let $0<\theta<1$, $1<p<\infty$
Let $N$ be a number with $0\le N<\infty$, $N^2+\inf_{0<t<T_0} M^2\ge0$.
Let $w_0$ and $w_1$ be real numbers with 
\beq
\label{w0}
w_0>\sup_{0<t<T_0} 
\left[
e^{-cNt} 
\left\{
\frac{N^2+M^2(t)}{(1-\theta)b(t)}
\right\}^{1/(p-1)}
\right],
\ \ 
w_1\ge cN w_0.
\eeq
Then the solution $w$ of the Cauchy problem 
\beq
\label{Cauchy-w}
\begin{cases}
c^{-2} \ddot{w}(t)+M^2(t)w(t)-b(t)|w|^p(t)\ge 0, 
\\
w(0)=w_0,\ \ \dot{w}(0)=w_1
\end{cases}
\eeq
for $0\le t<T_0$ satisfies the followings, 
where $\dot{w}:=dw/dt$ and $\ddot{w}:=d^2w/dt^2$.

(1) $w(t)\ge w_0 e^{cNt}$.

(2) $(1-\theta)b w^{p-1}-M^2>N^2$.

(3) $c^{-2} \ddot{w}-N^2w-\theta bw^p\ge0$.

(4) $\dot{w}\ge w_1$.
\end{lemma}

\vspace{10pt}

Taking the real part of the differential equation in \eqref{Cauchy}, 
and integrating by spatial variables, 
we have 
\[
c^{-2} \ddot{w}+M^2w-h=0,
\]
where $w(t):=\Re\int_\brn u(t,x)dx$ and 
$h(t):=\lambda a^{-n(p-1)/2}\int_\brn |u(t,x)|^pdx$.
We have 
\[
|w|^p\le (\omega_n r^n)^{p-1} \int_\brn |u(t,x)|^pdx
\]
by the finite speed of propagation of waves.
Thus, we obtain 
$h\ge b|w|^p$, where 
\[
b:=\lambda 
\left(
\omega_n^{2/n} a r^2
\right)^{-n(p-1)/2}>0
\]
since $\lambda>0$, and $a$, $r$ are positive valued.
So that, we obtain 
\[
c^{-2} \ddot{w}+M^2w-b|w|^p\ge 0,
\]
and 
the results in Lemma \ref{Lem-21} hold 
under the assumptions there, 
which are assumed in Theorem \ref{Thm-2}.

We use the test function method based 
in \cite{Nishihara-2012-OsakaJMath, QiS.Zhang-2001}  
for a semilinear damped wave equation in the Minkowski spacetime.
We extend it to the Klein-Gordon equation \eqref{Cauchy} in FLRW spacetimes 
(see \cite{Nakamura-Sato-2021-KJM} for the complex Ginzburg-Landau type equation in FLRW spacetimes).

We define a nonnegative function $\eta\in C^{\infty}_0([0,\infty))$ by 
\beq
\label{Blow-Up-500}
\eta(t)
:=
\begin{cases}
1 & \mbox{if}\ \ 0\le t\le \frac{1}{2}, 
\\
1-\eta_0\int_{1/2}^t e^{-(s-1/2)^{-1}(1-s)^{-1} } ds
& \mbox{if}\ \ \frac{1}{2}<t<1, 
\\
0 & \mbox{if}\ \ t\ge1, 
\end{cases}
\eeq
where $\eta_0:=\left(\int_{1/2}^1 e^{-(s-1/2)^{-1}(1-s)^{-1} } ds\right)^{-1}$.
We have 
\beq
\frac{|\partial_t\eta|^2}{\eta}\in C_0(\br).
\label{blowup-1}
\eeq
Put 
\beq
\label{Def-phi}
\phi(x):=\eta(|x|)
\eeq 
for $x\in \brn$.
Then we have 
\[
\phi(x)
=
\begin{cases}
1 \quad&\mbox{if} \ |x|\le\frac{1}{2}, 
\\ 
0&\mbox{if} \ |x|\ge1
\end{cases} 
\]
and 
\[
\frac{|\nabla \phi|^2}{\phi}
\left(=\frac{|\eta'|^2}{\eta}\right) \in C_0(\brn).
\]
For $R>0$, we put 
\beq
\label{Def-Eta-R}
\eta_R(t):=\eta\left(\frac{t}{R}\right), 
\ \ 
\phi_R(x):=\phi\left(\frac{x}{R}\right),
\ \ 
\psi_R(t,x):=\eta_R(t)\phi_R(x)
\eeq
for $0\le t<\infty$ and $x\in \brn$.
Then for $1<p\le \infty$, 
we have 
\beq
\label{Blow-Up-1000-C}
\psi_R^{p'}\in C_0^2(\br^{1+n}),
\eeq  
\beq
\label{Blow-Up-1000-t}
|\partial_t^2\psi_R^{p'}(t,x)|
\lesssim
\frac{1}{R^2}
\ \chi_{ \{s\ge0;\ R/2\le s\le R\} }(t)
\ \chi_{ \{y\in \brn;\ |y|\le R\} }(x)
\psi_R^{p'-1}
\eeq
and
\beq
\label{Blow-Up-1000-x}
|\Delta \psi_R^{p'}(t,x)|
\lesssim
\frac{1}{R^2}
\ \chi_{\{s\ge0;\, s\le R\}}
\ \chi_{ \{y\in \brn;\ R/2\le |y|\le R\} }(x)
\psi_R^{p'-1}
\eeq
for $0\le t<\infty$ and $x\in \brn$, 
where $\chi_S$ denotes the characteristic function on the set $S$.

Putting the test function $\psi$ in \eqref{Weak-Sol-Psi} 
as $\psi^{p'}_R$, 
and using $\partial_t \phi^{p'}_R(0)=0$, 
we have 
\beq
\label{Proof-Thm-2-1000}
\lambda I_R=-c^{-2}V_R+c^{-2}\two_R-\three_R+\four_R,
\eeq
where 
\begin{eqnarray*}
I_R 
&:=&
\int_0^\infty\int_\brn a^{-n(p-1)/2}|u(t,x)|^p\psi_R^{p'}(t,x) dx dt,
\\
\two_R 
&:=&
\int_0^\infty\int_\brn \Re u(t,x) \partial_t^2 \psi_R^{p'}(t,x) dxdt,
\\
\three_R 
&:=&
\int_0^\infty\int_\brn \Re u(t,x) a^{-2}(t)\Delta \psi_R^{p'}(t,x) dxdt,
\\
\four_R 
&:=&
\int_0^\infty\int_\brn \Re u(t,x) M^{2}(t) \psi_R^{p'}(t,x) dxdt,
\\
V_R 
&:=&
\int_\brn \Re u_1(x) \psi_R^{p'}(0,x) dx.
\end{eqnarray*}
We note $I_R\ge0$ by the definition. 
We estimate $\two_R$, $\three_R$, $\four_R$ and $V_R$ as follows.
We have 
\beq
\label{Proof-Thm-2-2000}
|\two_R|
\lesssim 
R^{-2}
\int_{R/2}^R\int_{|x|<R} |u(t,x)| \psi_R^{p'-1}(x) dx dt
\le
R^{-2}I_R^{1/p}\, \two_R'^{1/p'}
\eeq
by \eqref{Blow-Up-1000-t} and the H\"older inequality, 
where 
\beq
\label{II'}
\two_R':=\omega_n\int_{R/2}^R \left(R\wedge r(t)\right)^n a^{n/2}(t) dt,
\eeq
$R\wedge r(t):=\min\{R,r(t)\}$, 
and $\omega_n$ denotes the volume of unit ball in $\brn$.
Similarly, 
we have 
\beq
\label{Proof-Thm-2-3000}
|\three_R|
\lesssim 
R^{-2}
\int_{0}^R\int_{R/2<|x|<R} a^{-2}(t) |u(t,x)| \psi_R^{p'-1}(x) dx dt
\le
R^{-2}I_R^{1/p}\, \three_R'^{1/p'}
\eeq
by \eqref{Blow-Up-1000-x} and the H\"older inequality, 
where 
\beq
\label{III'}
\three_R':=\int_0^R a^{n/2-2p'} \int_{R/2<|x|<R\wedge r(t)} dx.
\eeq
We have $V_R=w_1$ by $\psi_R(0,\cdot)=1$.
We note $\four_R=0$ when $M^2\equiv0$.
When $M^2\not\equiv0$, we have 
\beq
\label{Proof-Thm-2-4000}
\four_R
\to \int_0^{T_0} \int_\brn (\Re u) M^2(t) dx dt
=\int_0^{T_0} M^2(t) w (t) dt<0
\eeq
by the Lebesgue convergence theorem, 
$0\le \psi_R\le 1$,
$M^2\le 0$, and $w(t)\ge w_0>0$ for $t\ge0$. 
Thus, we obtain $\four_R\le 0$ for sufficiently large $R$.
By \eqref{Proof-Thm-2-1000}, \eqref{Proof-Thm-2-2000}, 
\eqref{Proof-Thm-2-3000} and \eqref{Proof-Thm-2-4000},
we have 
\[
0<\lambda I_R^{1/p'}\lesssim R^{-2}
\left(\two_R'^{1/p'}+\three_R'^{1/p'}\right).
\]
Thus, we have 
\beq
\label{Proof-Thm-2-5000}
\lim_{R\to\infty} I_R=0
\eeq 
under 
\[
\lim_{R\to\infty} R^{-2} \two_R'^{1/p'} 
=
\lim_{R\to\infty} R^{-2} \three_R'^{1/p'} 
=0
\]
which are the assumptions 
\eqref{Thm-2-Assume-1} 
and 
\eqref{Thm-2-Assume-2}.
Since we have 
\begin{eqnarray*}
I_R
&=& 
\int_0^{\infty} \int_\brn a^{-n(p-1)/2} |u|^p(t,x) \psi_R^{p'}(t,x) dx dt 
\\
&\ge& 
\int_0^{R/2} \int_{|x|<R/2}  a^{-n(p-1)/2} |u|^p(t,x) dx dt
\end{eqnarray*}
for any $0<R<\infty$,
we obtain $u=0$ on $[0,\infty)\times\brn$ by \eqref{Proof-Thm-2-5000}.

\newsection{Proof of Corollary \ref{Cor-3}}
\label{Sec-Proof-Cor-3}
We prove Cororally \ref{Cor-3} in this section.
We note that $T_0=\infty$ in \eqref{R-Def-T_0} holds if and only if 
$H=0$ with $\sigma\in\br$, $H\neq0$ with $\sigma=-1$, 
$H>0$ with $\sigma>-1$, and $H<0$ with $\sigma<-1$.
We use the following fundamental results.

\begin{lemma}
([\cite[Lemmas 2.6 and 2.7]{Nakamura-Yoshizumi-2025-JDE}].)
\label{Lem-11}
Let $m\in \br$, $H\in \br$ and $\sigma\in \br$.
Let $T_0$, $a$ and $M$ be given by \eqref{R-Def-T_0}, \eqref{Def-a} and \eqref{M-FLRW}.
Then the following results hold.

(1) 
\ $\frac{\dot{a}}{{a}}=H\left(\frac{a}{a_0}\right)^{-n(1+\sigma)/2}$.

(2) 
\ $\displaystyle M^2
=
m^2+\sigma\left(\frac{nH}{2c}\right)^2
\left\{1+\frac{n(1+\sigma)Ht}{2}\right\}^{-2}$.

(3) 
\ $M^2$ satisfies 
\[
\begin{array}{ll}
(i) \ \ M^2=m^2 & \mbox{if }\ H=0, \ \mbox{or}\ \sigma=0,
\\
(ii)\ \ M^2=m^2-\left(\frac{nH}{2c}\right)^2 & \mbox{if }\ H\neq0, \ \sigma=-1,
\\
(iii)\ \ m^2<M^2\le m^2+\sigma\left(\frac{nH}{2c}\right)^2 & \mbox{if }\ H>0, \ \sigma>0,
\\
(iv)\ \ m^2+\sigma\left(\frac{nH}{2c}\right)^2\le M^2<m^2 & \mbox{if }\ (1+\sigma)H>0, \ \sigma<0,
\\
(v)\ \ m^2+\sigma\left(\frac{nH}{2c}\right)^2\le M^2\to \infty \ (t\to T_0) & \mbox{if }\ H<0, \ \sigma>0,
\\
(vi)\ \ m^2+\sigma\left(\frac{nH}{2c}\right)^2\ge M^2\to -\infty \ (t\to T_0) & \mbox{if }\ (1+\sigma)H<0, \ \sigma<0.
\end{array}
\]
\end{lemma}

By (v) and (vi) in (3) in Lemma \ref{Lem-11}, 
the cases $H<0$ with $\sigma>0$, $H>0$ with $\sigma<-1$, 
and $H<0$ with $-1<\sigma<0$ are eliminated under the restrictions 
\beq
\label{Proof-Cor-3-1000}
M^2\le0,\ \ \ \ N^2+M^2\ge0
\eeq
for some constant $N\ge0$, 
which are assumed in Theorem \ref{Thm-2} 
as $M\in C([0,\infty),i\br)$ and \eqref{Thm-2-Assume-aMN}.

By Lemma \ref{Lem-11}, we have 

\begin{tabular}{ll}
$M^2=m^2$ 
& if  $H=0$ with $\sigma\in \br$, or $H>0$ with $\sigma=0$,
\\
$M^2=m^2-(nH/2c)^2$ 
& if  $H>0$ with $\sigma=-1$, or $H<0$ with $\sigma=-1$,
\\
$m^2<M^2\le m^2+\sigma\left(\frac{nH}{2c}\right)^2$ 
& if  $H>0$ with $\sigma>0$, 
\\
$m^2+\sigma\left(\frac{nH}{2c}\right)^2\le M^2<m^2$ 
& if  $H>0$ with $\sigma>-1$, or $H<0$ with $\sigma<-1$.
\end{tabular}

We have 
\beq
\label{r}
r=r_0+
\begin{cases}
\frac{ct}{a_0} & \mbox{if}\ H=0\ \mbox{with}\ \sigma \in \br,
\\
\frac{c}{a_0H}\left(1-e^{-Ht}\right) 
& \mbox{if}\ H>0\ \mbox{with}\ \sigma=-1,
\\
-\frac{c}{a_0H}\left(e^{-Ht}-1\right) 
& \mbox{if}\ H<0\ \mbox{with}\ \sigma=-1,
\\
\frac{2c}{ a_0H\left\{n(1+\sigma)-2\right\} }
\left\{
\left(
\frac{a}{a_0}
\right)^{n(1+\sigma)/2-1}
-1
\right\}
& \mbox{if}\ H>0\ \mbox{with}\ \sigma \neq -1+\frac{2}{n},
\\
& \mbox{or}\  H<0\ \mbox{with}\ \sigma <-1,
\\
\frac{2c}{ a_0H\left\{2-n(1+\sigma)\right\} }
\left\{
1-
\left(
\frac{a}{a_0}
\right)^{n(1+\sigma)/2-1}
\right\}
& \mbox{if}\ H>0\ \mbox{with}\ -1<\sigma<-1+\frac{2}{n},
\\
\frac{c}{a_0H} \log(1+Ht)
& \mbox{if}\ H>0\ \mbox{with}\ \sigma = -1+\frac{2}{n}.
\end{cases}
\eeq
We recall the definitions of $\two'_R$ and $\three'_R$ in 
\eqref{II'} and \eqref{III'}.

(1) 
We consider the case $H=0$ with $\sigma\in \br$.
We have $M^2=m^2\le 0$ and $N^2+M^2\ge0$ if $m\in i\br$ and $N=|m|$.
Thus, \eqref{Thm-2-Assume-aMN} holds.
We also have $S=0$ in \eqref{Def-b-S}.
We have 
$\two'_R\lesssim R^{n+1}$ and $\three'_R\lesssim R^{n+1}$ 
by $R\wedge r(\cdot)\le R$ and $a(\cdot)=a_0$.
Thus, the conditions \eqref{Thm-2-Assume-1} and \eqref{Thm-2-Assume-2} 
hold if $p<(n+1)/(n-1)$.

(2) 
We consider the case $H>0$ with $\sigma>-1$.
Since $a(\cdot)$ and $r(\cdot)$ are polynomial order at most 
by \eqref{Def-a} and \eqref{r}, 
the function $b(\cdot)$ is also polynomial order by its definition in \eqref{Def-b-S}.
We consider the following three cases.

(I) 
We consider the case $H>0$ with $\sigma>0$.
We have $m^2<M^2\le m^2+\sigma(nH/2c)^2\le 0$ 
and 
$N^2+\inf M^2=N^2+m^2=0$ if $m\in i\br$ and $N=|m|$.
Thus, \eqref{Thm-2-Assume-aMN} holds.
We also have 
$\lim_{t\to\infty}S(t)=0$ 
in \eqref{Def-b-S}.
Especially, $\sup_{0<t<\infty} S(t)<\infty$.

(II) 
We consider the case $H>0$ with $\sigma=0$.
We have $M^2=m^2\le 0$ and $N^2+M^2=0$ if $m\in i\br$ and $N=|m|$.
Thus, \eqref{Thm-2-Assume-aMN} holds.
We also have $S=0$ in \eqref{Def-b-S}.
Especially, $\sup_{0<t<\infty} S(t)<\infty$.

(III) 
We consider the case $H>0$ with $-1<\sigma<0$.
We have 
$m^2+\sigma(nH/2c)^2\le M^2<m^2<0$ and 
$N^2+\inf_{0<t<\infty} M^2=N^2+m^2+\sigma(nH/2c)^2=0$ 
if $m\in i\br$ and $N=\sqrt{-m^2-\sigma(nH/2c)^2}$.
Thus, \eqref{Thm-2-Assume-aMN} holds.
We also have $\lim_{t\to\infty}S(t)=0$ in \eqref{Def-b-S} 
since $M^2$ is bounded, $b$ is polynomial order, and $N>0$.
Especially, $\sup_{0<t<\infty} S(t)<\infty$.

Under the above cases (I), (II) and (III), 
we check the conditions 
\eqref{Thm-2-Assume-1} 
and 
\eqref{Thm-2-Assume-2}.
Since $r$ is increasing function, the condition $R/2\le r(t)$ leads $R/2\le r(R)$ for $0\le t\le R$.
However, since $r(R)$ behaves like 
\beq
\label{rR}
r(R)\simeq 
\begin{cases}
R^{1-2/n(1+\sigma)} & \mbox{if}\ \ \sigma>-1+2/n, \\
\log R & \mbox{if}\ \ \sigma=-1+2/n,\\ 
1 & \mbox{if}\ \ \sigma<-1+2/n,
\end{cases}
\eeq
the inequality $R/2\le r(R)$ does not hold for sufficiently large $R$.
Thus, we have $\three'_R=0$, and $R^{-2}\three_R'^{1/p'}=0$ for sufficiently large $R$, 
which yield \eqref{Thm-2-Assume-2}.
Since $r$ and $a$ are increasing functions, 
we have $R\wedge r(t)\le r(t)\le r(R)$ and $a^{n/2}(t)\le a^{n/2}(R)$ 
for $R/2\le t\le R$.
Thus, we have 
\[
\two_R'\le \frac{\omega_n Rr(R)^n a^{n/2}(R)}{2}.
\]
By \eqref{rR} and $a(R)\lesssim R^{2/n(1+\sigma)}$ for sufficiently large $R>0$, 
we obtain 
\beq
\two_R'
\lesssim
\begin{cases}
R^{1+n-1/(1+\sigma)} & \mbox{if}\ \ \sigma>-1+2/n, \\
R^{1+1/(1+\sigma)}(\log R)^n & \mbox{if}\ \ \sigma=-1+2/n,\\ 
R^{1+1/(1+\sigma)} & \mbox{if}\ \ \sigma<-1+2/n
\end{cases}
\eeq
for sufficiently large $R$.
So that, we have 
\[
\lim_{R\to\infty} R^{-2}\two_R'^{1/p'}=0
\]
under one of the three conditions given by  
\begin{eqnarray}
&&-2+\left(1+n-\frac{1}{1+\sigma}\right)\frac{1}{p'}<0 
\ \ \mbox{with} \ \ \sigma>-1+\frac{2}{n}, \label{psigma-1}\\
&&-2+\left(1+\frac{1}{1+\sigma}\right)\frac{1}{p'}<0 
\ \  \mbox{with} \ \ \sigma=-1+\frac{2}{n}, \label{psigma-2}\\
&&-2+\left(1+\frac{1}{1+\sigma}\right)\frac{1}{p'}<0 
\ \ \mbox{with} \ \ \sigma<-1+\frac{2}{n}.\label{psigma-3}
\end{eqnarray}
Since we have 
$(1+n)(1+\sigma)-1>1+2/n>0$ for $\sigma>-1+2/n$,
the condition \eqref{psigma-1} is satisfied if 
\beq
\label{psigma-1-p}
1<p<
\begin{cases}
\infty & \mbox{if}\ \ n=1,
\\
\frac{(n+1)(1+\sigma)-1}{(n-1)(1+\sigma)-1} & \mbox{if}\ \ n\ge2
\end{cases}
\eeq
with $\sigma>-1+2/n$,
where we note $(n-1)(1+\sigma)-1>1-2/n\ge0$ when $n\ge2$.
The condition \eqref{psigma-2} is rewritten as 
\beq
\label{psigma-2-p}
1<p<
\begin{cases}
\infty & \mbox{if}\ \ n=1,2, 
\\
\frac{n+2}{n-2} & \mbox{if}\ \ n\ge3
\end{cases}
\eeq
with $\sigma=-1+2/n$.
The condition \eqref{psigma-3} is rewritten as 
\beq
\label{psigma-3-p}
1<p<
\begin{cases}
\infty & \mbox{if}\ \ \sigma\ge0, 
\\
-\frac{2+\sigma}{\sigma} & \mbox{if}\ \ -1<\sigma<0
\end{cases}
\eeq
with $\sigma<-1+2/n$, 
where we note $-(2+\sigma)/\sigma>1$ when $\sigma>-1$.

The conditions \eqref{psigma-1-p}, \eqref{psigma-2-p} and \eqref{psigma-3-p} are rewritten as \eqref{Cor-3-2-p}, 
by which the condition \eqref{Thm-2-Assume-1} is satisfied.

(3) 
We consider the case $H<0$ and $\sigma<-1-2/n$.
In this case, $a(\cdot)$ is monotone decreasing by its definition \eqref{Def-a}, and 
$r(t)\to\infty$ as 
$t\to\infty$ by \eqref{r}.
We have $m^2+\sigma(nH/2c)^2\le M^2<m^2\le 0$ 
and 
$N^2+\inf M^2=N^2+m^2+\sigma(nH/2c)^2=0$ if $m\in i\br$ 
and $N=\sqrt{-m^2-\sigma(nH/2c)^2}$.
We note $N>0$ by $H<0$ and $\sigma<-1$.
Since $a(\cdot)$ and $r(\cdot)$ are polynomial order at most, 
the function $b(\cdot)$ is also polynomial order by its definition in \eqref{Def-b-S}.
Thus, $\lim_{t\to\infty} S(t)=0$.
Especially, $\sup_{0<t<\infty} S(t)<\infty$.
Let $t_R$ be the number with $R/2=r(t_R)$.
Since 
\[
r(t_R)=r_0
-
\frac{2c}{ a_0H\left\{2-n(1+\sigma)\right\} }
\left\{
\left(
1+\frac{n(1+\sigma)Ht_R}{2}
\right)^{1-2/n(1+\sigma)}
-1
\right\},
\]
we have 
\beq
\label{Proof-Cor-3-5-tR}
t_R\simeq R^{n(1+\sigma)/\{n(1+\sigma)-2\} }
\eeq
for sufficiently large $R$.
Thus,
$t_R\lesssim R$ holds for sufficiently large $R$ 
since $0<n(1+\sigma)/\{n(1+\sigma)-2\}<1$ holds for $\sigma<-1$.

We estimate $\three_R'$.
We have 
\[
\three_R'\le \iint_{\substack{t_R<t<R \\ R/2<|x|<R}} a^{n/2-2p'}(t) dx dt
\]
since $R/2<|x|<r(t)$ in the definition \eqref{III'} requires $t\ge t_R$.
Thus, we have 
\[
\three_R'
\lesssim 
R^n\cdot
\begin{cases}
R^{\gamma+1} \vee t_R^{\gamma+1} 
& \mbox{if}\ \ \gamma\neq-1,
\\
\log\left\{ 1+\frac{n(1+\sigma)HR}{2}\right\} 
& \mbox{if}\ \ \gamma=-1,
\end{cases}
\]
where we have put $\gamma:=(1-4p'/n)/(1+\sigma)$.
By \eqref{Proof-Cor-3-5-tR} and $t_R< R$, 
we have 
\[
R^{-2} \three_R'
\lesssim
\begin{cases}
R^{-2+(n+\gamma+1)/p'} & \mbox{if}\ \ \gamma>-1,
\\
R^{-2+[n+n(1+\sigma)(\gamma+1)/\{n(1+\sigma)-2\}]/p'} & \mbox{if}\ \ \gamma<-1,
\\
R^{-2+n/p'} \left(\log R\right)^{1/p'} & \mbox{if}\ \ \gamma=-1.
\end{cases}
\]
So that, the condition \eqref{Thm-2-Assume-1} holds if one of the following three conditions holds; 
\beq
\label{Proof-Cor-3-5-Gamma}
\begin{cases}
(A1)\ \ \gamma>-1 \ \ \mbox{with}\ \ (A2)\ \ -2+\frac{n+\gamma+1}{p'}<0, 
\\
(A3)\ \ \gamma<-1 \ \ \mbox{with}\ \ 
(A4)\ \ -2+\frac{n+n(1+\sigma)(\gamma+1)/\{n(1+\sigma)-2\}}{p'}<0, 
\\
(A5)\ \ \gamma=-1 \ \ \mbox{with}\ \ (A6)\ \ -2+\frac{n}{p'}<0,
\end{cases}
\eeq
where we have put the numbers from $(A1)$ to $(A6)$.
We also have 
\[
\two_R'\le \omega_n a_0^{2p'}R^n\int_{t_R}^R a^{n/2-2p'}(t)dt
\]
by $t_R\le R/2$, $R\wedge r(\cdot)\le R$ and $a^{n/2}\le a^{n/2-2p'}a_0^{2p'}$,
where we note that $a$ is a decreasing function.
Thus, we obtain \eqref{Thm-2-Assume-2} under \eqref{Proof-Cor-3-5-Gamma}.
Now, we consider the conditions \eqref{Proof-Cor-3-5-Gamma}.

Let us consider the condition (A1), 
which is rewritten as 
$p'>n(2+\sigma)/4$ under $\sigma<-1$.
The condition $-2+{(n+\gamma+1)}/{p'}<0$ is rewritten as 
\[
(n+1)(1+\sigma)+1>\frac{2p'\left\{2+n(1+\sigma)\right\}}{n}
\]
under $\sigma<-1$, 
which is equivalent to one of the following three conditions given by 
\beq
\label{Proof-Cor-3-5-p'}
\begin{cases}
(B1)\ \ p'<\frac{n\left\{(n+1)(1+\sigma)+1\right\}}{2\left\{2+n(1+\sigma)\right\}}
& \mbox{with} \ \ (B2)\ \ 2+n(1+\sigma)>0,
\\
(B3)\ \ (n+1)(1+\sigma)+1>0 
& \mbox{with} \ \ (B4)\ \ 2+n(1+\sigma)=0,
\\
(B5)\ \ p'>\frac{n\left\{(n+1)(1+\sigma)+1\right\}}{2\left\{2+n(1+\sigma)\right\}}
& \mbox{with} \ \ (B6)\ \ 2+n(1+\sigma)<0,
\end{cases}
\eeq
where we have put the numbers from $(B1)$ to $(B6)$.
If $2+n(1+\sigma)>0$, then 
(A1) and (B1) are rewritten as 
\[
\frac{n(2+\sigma)}{4}
<p'<
\frac{n\left\{(n+1)(1+\sigma)+1\right\}}{2\left\{2+n(1+\sigma)\right\}},
\]
which requires 
\[
2(1+\sigma)+n(2+\sigma)(1+\sigma)<2(n+1)(1+\sigma).
\]
Under $\sigma<-1$, this is rewritten as $\sigma>0$ which is a contradiction to $\sigma<-1$.
So that, the first case $(B1)$ under $(B2)$ in \eqref{Proof-Cor-3-5-p'} does not occur under (A1).
The second case $(B3)$ under $(B4)$ shows $\sigma>-1-1/(n+1)$ if $\sigma=-1-2/n$, which is a contradiction.
Let us consider the case $(B5)$ under $(B6)$ as follows.
We note that $(B6)$ is rewritten as $\sigma<-1-2/n$.

We compare (A1) and (B5).
Since 
\[
\frac{n(2+\sigma)}{4}<
\frac{n\left\{(n+1)(1+\sigma)+1\right\}}{2\left\{2+n(1+\sigma)\right\}}
\]
is rewritten as 
\beq
\label{Proof-Cor-3-A1B5}
2(n+1)(1+\sigma)<\left\{2+n(2+\sigma)\right\}(1+\sigma)
\eeq
under (B6),  
and \eqref{Proof-Cor-3-A1B5} is rewritten as $\sigma<0$ 
under $\sigma<-1$,
the condition (B5) under (B6) implies the condition (A1).

We compare $p'>1$ and (B5).
Since 
\beq
\label{Proof-Cor-3-p'1B5}
1<
\frac{n\left\{(n+1)(1+\sigma)+1\right\}}{2\left\{2+n(1+\sigma)\right\}}
\eeq
is rewritten as 
\beq
\label{Proof-Cor-3-p'1B5'}
n(n-1)(1+\sigma)<4-n
\eeq
under (B6),  
and \eqref{Proof-Cor-3-p'1B5'} is rewritten as 
\[
\sigma
<\begin{cases}
-1 & \mbox{if}\ \ n\le 4,
\\
-1-\frac{n-4}{n(n-1)} & \mbox{if}\ \ n\ge5,
\end{cases}
\]
the inequality \eqref{Proof-Cor-3-p'1B5} holds under (B6),
where we note that 
(B6) yields $\sigma<-1-(n-4)/n(n-1)$.
Thus, (B5) yields $p'>1$ under (B6).
We note that (B5) is rewritten as 
\[
\frac{1}{p}
>1-\frac{ 2\{2+n(1+\sigma)\} }{ n\{(n+1)(1+\sigma)+1\}},
\]
namely, $p<p_0$ under (B6), 
where $p_0$ is defined by \eqref{p0} and $1<p_0<\infty$ holds 
since  $2\{2+n(1+\sigma)\}/n\{(n+1)(1+\sigma)+1\}<1$ holds 
by \eqref{Proof-Cor-3-p'1B5}.
We have 
$p_0=1$ under (B4). 
We also have 
\[
p_0\to 
\begin{cases}
\frac{n(n+1)}{n(n-1)} & \mbox{if}\ \ n\ge2,
\\
\infty & \mbox{if}\ \ n=1
\end{cases}
\]
as $\sigma\to-\infty$.
Since (B6) implies $n(n-1)(1+\sigma)+n-4<0$, 
we have 
\[
p_0<\frac{n+1}{n-1}
\]
when $n\ge2$.
So that,
the condition (A1) under (A2) is equivalent to the condition (B5) under (B6),
namely, to the condition 
\beq
\label{Proof-Cor-3-p0}
1<p<p_0 \ \ \mbox{and}\ \ 
\sigma<-1-\frac{2}{n}.
\eeq

The condition (A3) is rewritten as 
$p'<n(2+\sigma)/4$ under $\sigma<-1$.
Since (A4) is rewritten as 
\[
n(1+\sigma)(\gamma+1)>(2p'-n)\{n(1+\sigma)-2\}
\]
by $n(1+\sigma)-2<0$ under $\sigma<-1$, 
(A4) is rewritten as $p'>(n+1)/2-1/2(1+\sigma)$. 
Thus, (A3) with (A4) is equivalent to 
\[
\frac{n+1}{2}-\frac{1}{2(1+\sigma)}<p'<\frac{n(2+\sigma)}{4}.
\]
However, this inequality requires $(2-n\sigma)(1+\sigma)>2$, 
which does not hold under $\sigma<-1$.
So that, (A3) with (A4) does not hold.

The condition (A5) is rewritten as 
$p'=n(2+\sigma)/4$, 
and (A6) is rewritten as $\sigma>0$ under (A5), 
which contradicts to $\sigma<-1$.
So that, (A5) with (A6) does not hold.
Therefore, the three conditions in \eqref{Proof-Cor-3-5-Gamma} reduce to the first condition (A1) with (A2).
The condition (A1) and (A2) is rewritten as \eqref{Proof-Cor-3-p0}, 
which is the condition appeared in \eqref{p0}. 
We have proved Corollary \ref{Cor-3}.

\vspace{10pt}

\begin{remark}
\label{Rem-4}
We remark why the argument to prove Corollary \ref{Cor-3} does not work well 
except for the conditions (1), (2) and (3) in the corollary.
As a typical example, 
we only refer to the case of the de Sitter spacetime, namely, $H\neq0$ and $\sigma=-1$ in \eqref{Def-a}.
In this case, we have $M^2=m^2-(nH/2c)^2$ by \eqref{M-FLRW}, 
and the required condition $M^2\le 0$ in Theorem \ref{Thm-2} holds 
if and only if 
$m\in \br$ with $|m|\le n|H|/2c$, or $m\in i\br$.
We have $N^2+M^2=0$ and $S=0$ putting $N:=\sqrt{(nH/2c)^2-m^2}$, 
by which the conditions in \eqref{Thm-2-Assume-aMN} hold.
So that, the conditions in \eqref{Assume-S-w} hold, provided $w_0>0$ and $w_1\ge cNw_0$.
Let us check the conditions \eqref{Thm-2-Assume-1} and \eqref{Thm-2-Assume-2}.

Firstly, we consider the case $H>0$.
Since $r$ is bounded from above as $r\le r_0+c/a_0H$ by \eqref{r}, 
we have 
\[
\two_R'=
\left(
r_0+\frac{c}{a_0H}
\right)^n \frac{a_0^n e^{nHR}(1-e^{-nHR/2})}{nH}
\]
for sufficiently large $R$,
where $\two_R'$ is defined by \eqref{II'}.
So that, we have $\lim_{R\to\infty} R^{-2}$ $\two_R'^{1/p'}\to \infty$,
which shows the condition \eqref{Thm-2-Assume-1} does not hold.
On the other hand, 
we have $\three_R'=0$ since the integral region in \eqref{III'} under $R/2<R\wedge r$ is empty for sufficiently large $R$.
So that, the condition \eqref{Thm-2-Assume-2} holds.

Secondly, we consider the case $H<0$.
Since $r(t)$ behaves like $e^{|H|t}$ by \eqref{r}, 
we have $R\wedge r(t)=R$ for sufficiently large $R$.
So that, we have 
\[
\two_R'\lesssim R^n e^{nHR/4},
\]
which yields $\lim_{R\to\infty} R^{-2} \two_R'^{1/p'}=0$.
Namely, the condition \eqref{Thm-2-Assume-1} holds.
On the other hand, 
we have 
\beq
\three_R'
=
\omega_n
a_0^{n/2-2p'}
\int_{R/2}^R\int_{t(r)}^R e^{(n/2-2p')Ht} dt r^{n-1} dr,
\eeq
where $t(r)$ is the inverse function of \eqref{r} defined by 
\[
t(r):=-\frac{1}{H}\log\left(1-\frac{a_0H(r-r_0)}{c}\right).
\]
When $n/2-2p'>0$, 
we have $\three_R'\simeq R^{n/2+2p'}$ for sufficiently large $R$,
and the condition $\lim_{R\to\infty}R^{-2}\three_R'^{1/p'}=0$ requires $n/2p'<0$, which contradicts to $1<p<\infty$.
When $n/2-2p'<0$, 
we have $\three_R'\simeq e^{(n/2-2p')HR}R^n$ for sufficiently large $R$,
and $\lim_{R\to\infty}R^{-2}\three_R'^{1/p'}=\infty$.
When $n/2-2p'=0$, 
we have $\three_R'\simeq R^{n+1}$ for sufficiently large $R$,
and $\lim_{R\to\infty}R^{-2}\three_R'^{1/p'}=\infty$.
So that, the condition \eqref{Thm-2-Assume-2} does not hold.
\end{remark}

%%%%%%%%%%%%%%%%%%%%%%%%%%%%%%%%%%
%
%%%%%%%%%%%%%%%%%%%%%%%%%%%%%%%%%%

{\bf Acknowledgments.}
This work was supported by JSPS KAKENHI Grant Numbers 16H03940, 17KK0082, 22K18671.
%The authors are thankful to the anonymous referee for several comments to revise the paper.
%The author is thankful to the anonymous referee for several comments to revise the paper.

%{\bf Data Availability.}
%Data sharing is not applicable to this article as no new data were created or analyzed in this study.

%%%%%%%%%%%%%%%%%%%%%%%%%%%%%%%%%%
%
%%%%%%%%%%%%%%%%%%%%%%%%%%%%%%%%%%


\begin{thebibliography}{999}

\if0%%%%%
\bibitem{Balogh-Banda-Yagdjian-2019-CommNonlinearSciNumerSimulat} 
A. Balogh, J. Banda, K. Yagdjian, 
%Balogh, Andras(1-UTRGV); Banda, Jacob(1-UTRGV); Yagdjian, Karen(1-UTRGV)
\emph{High-performance implementation of a Runge-Kutta finite-difference scheme for the Higgs boson equation in the de Sitter spacetime}, 
Commun. Nonlinear Sci. Numer. Simul. {\bf 68} (2019), 15--30. 
\fi %%%%%%

\bibitem{Baskin-2012-AHP}
D. Baskin,
%Dean Baskin
\emph{Strichartz estimates on asymptotically de Sitter spaces},
Annales Henri Poincar{\'e} {\bf 14} (2013), Issue 2, pp 221--252.

\bibitem{Carroll-2004-Addison}
S. Carroll, 
%Carroll, Sean(1-CHI)
\emph{Spacetime and geometry.  
An introduction to general relativity}, 
Addison Wesley, San Francisco, CA, 2004, xiv+513 pp.

\bibitem{DInverno-1992-Oxford}
R. d'Inverno, 
%d'Inverno, Ray(4-SHMP)
\emph{Introducing Einstein's relativity}, 
The Clarendon Press, Oxford University Press, New York, 1992, xii+383 pp. 

\bibitem{Galstian-Yagdjian-2015-NA-TMA} 
A. Galstian, K. Yagdjian, 
%Galstian, Anahit(1-UTRGV-M); Yagdjian, Karen(1-UTRGV-M)
\emph{Global solutions for semilinear Klein-Gordon equations in FLRW spacetimes},
%(English summary)
Nonlinear Anal. {\bf 113}(2015), 339--356.

\bibitem{Galstian-Yagdjian-2020-RMP}
A. Galstian, K. Yagdjian, 
%Galstian, Anahit; Yagdjian, Karen
\emph{Finite lifespan of solutions of the semilinear wave equation in the Einstein-de Sitter spacetime}, 
Rev. Math. Phys. {\bf 32} (2020), no. 7, 2050018, 31 pp.

\bibitem{Hintz-Vasy-2015-AnalysisPDE}
P. Hintz, A. Vasy,  
%Hintz, Peter; Vasy, Andr\'as 
\emph{Semilinear wave equations on asymptotically de Sitter, Kerr-de Sitter and Minkowski spacetimes}, 
Anal. PDE  {\bf 8} (2015),  no. 8, 1807--1890.

\bibitem{Kato-1980-CPAM}
T. Kato, 
%Kato, Tosio
\emph{Blow-up of solutions of some nonlinear hyperbolic equations}, 
Comm. Pure Appl. Math.  {\bf 33} (1980),  no. 4, 501--505. 

\bibitem{McCollum-Mwamba-Oliver-2024-NA}
J. McCollum, G. Mwamba, J. Oliver, 
%McCollum, Jonathon(1-ORS); Mwamba, Gregory(1-CA2-AM); Oliver, Jesús(1-CSUEB-M)
\emph{A sufficient condition for blowup of the nonlinear Klein-Gordon equation with positive initial energy in FLRW spacetimes}, 
%(English summary)
Nonlinear Anal. {\bf 246} (2024), Paper No. 113582, 11 pp.

\bibitem{Nakamura-2014-JMAA}
M. Nakamura, 
%Nakamura, Makoto
\emph{The Cauchy problem for semi-linear Klein-Gordon equations in de Sitter spacetime},
J. Math. Anal. Appl.  {\bf 410}  (2014),  no. 1, 445--454. 

\bibitem{Nakamura-2021-JMP}
M. Nakamura, 
%Nakamura, Makoto(J-YAMATSC-NDM)
\emph{The Cauchy problem for the Klein-Gordon equation under the quartic potential in the de Sitter spacetime}, 
J. Math. Phys. {\bf 62} (2021), no. 12, Paper No. 121509, 21 pp.

\bibitem{Nakamura-Sato-2021-KJM}
M. Nakamura, Y. Sato,
%Nakamura, Makoto; Sato, Yuya
\emph{Existence and non-existence of global solutions for the semilinear complex Ginzburg-Landau type equation in homogeneous and isotropic spacetime},
Kyushu J. Math. 75 (2021), no. 2, 169–209.

\bibitem{Nakamura-Yoshizumi-2025-JMP}
M. Nakamura, T. Yoshizumi, 
\emph{Blowing-up solutions of  Klein-Gordon equations with gauge variant semilinear terms in Friedmann-Lema\^{i}tre-Robertson-Walker spacetimes under finite speed of propagation}, 
J. Math. Phys. {\bf 66} (2025), 041507. 
%Published Online: 8 April 2025.
%doi: 10.1063/5.025396

\bibitem{Nakamura-Yoshizumi-2025-JDE}
M. Nakamura, T. Yoshizumi,
{\it The Cauchy problem for semi-linear Klein-Gordon equations in Friedmann-Lema\^itre-Robertson-Walker spacetimes},
Journal of Differential Equations {\bf 438} (2025), 113395.
%Volume 438, 5 September 2025, 113395
%doi: 10.1016/j.jde.2025.113395

\bibitem{Nishihara-2012-OsakaJMath}
K. Nishihara,
\emph{Asymptotic behavior of solutions for a system of semilinear heat equations and the corresponding damped wave system},
Osaka J. Math. {\bf49} (2012), No. 2, 331--348.

\bibitem{QiS.Zhang-2001}
Q. Zhang, \emph{A blow-up result for a nonlinear wave equation with damping: The critical case}, 
C. R. Acad. Sci. Paris S\'{e}r. I Math. {\bf333} (2001), 109--114.

\if0%%%%%%%
\bibitem{Sogge-2008-IntPress}
C. D. Sogge, 
%Sogge, Christopher D.(1-JHOP)
\emph{Lectures on non-linear wave equations}, 
Second edition,
International Press, Boston, MA, 2008, x+205 pp.
\fi %%%%%%%

\bibitem{Wei-Yong-2024-JMP}
C. Wei, Z. Yong, 
%Wei, Changhua(PRC-ZSTU-M); Yong, Zikai(PRC-ZSTU-M)
\emph{Global existence and blowup of smooth solutions to the semilinear wave equations in FLRW spacetime}, 
%.(English summary)
J. Math. Phys. {\bf 65} (2024), no.5, Paper No. 051504, 21 pp.

\bibitem{Yagdjian-2009-DCDS}
K. Yagdjian, 
%Yagdjian, Karen 
\emph{The semilinear Klein-Gordon equation in de Sitter spacetime},
Discrete Contin. Dyn. Syst. Ser. S  {\bf 2} (2009),  no. 3, 679--696. 

\bibitem{Yagdjian-2012-JMAA}
K. Yagdjian,
%Yagdjian, Karen 
\emph{Global existence of the scalar field in de Sitter spacetime},
J. Math. Anal. Appl. {\bf 396} (2012), no. 1, 323--344. 

\end{thebibliography}
\end{document}